# Energy Inequality in Climate Hazards: Empirical Evidence of Social and Spatial Disparities in Managed and Hazard-Induced Power Outages

**Authors:** Natalie Coleman[1*], Amir Esmalian[2] , Cheng-Chun Lee[3] , Eulises Gonzales[4], Pranik Koirala[5] , Ali Mostafavi[6]

**Affiliations:**
[1] Ph.D. Student, Zachry Department of Civil and Environmental Engineering, Urban Resilience.AI Lab, Texas A&M University, College Station, TX, United States of America; email: ncoleman@tamu.edu
*Corresponding author
[2] Ph.D., Zachry Department of Civil and Environmental Engineering, Urban Resilience.AI Lab, Texas A&M University, College Station, TX, United States of America; email: amiresmalian@tamu.edu
[3] Postdoctoral research associate, Ph.D., Department of Civil and Environmental Engineering, Engineering, Texas A&M University, College Station, TX, United States of America; email: ccbarrylee@tamu.edu
[4] Undergraduate Researcher, Department of Computer Science and Engineering, Engineering, Texas A&M University, College Station, TX, United States of America; euli19_99@tamu.edu
[5] M.S. student, Department of Computer Science and Engineering, Engineering, Texas A&M University, College Station, TX, United States of America; email: pranik@tamu.edu
[6] Associate Professor, Zachry Department of Civil and Environmental Engineering, Urban Resilience.AI Lab, Texas A&M University, College Station, TX, United States of America; e-mail: amostafavi@civil.tamu.edu

## Abstract

Due to the effects of climate change and urbanization, the severity and frequency of hazard events is expected to increase. The energy sector in the United States  is ever more vulnerable to extreme climatic hazards: hurricane winds can damage electrical lines, causing hazard-induced power outages. Extreme heat and freezing temperatures can imbalance the supply and demand for energy resulting in managed power outages. Utility companies reportedly prioritize the restoration of power systems based on the number of outages and the size of affected populations. This approach fails to account for unequal impacts of hazard-induced and managed power outages. Research in equitable infrastructure emphasizes that certain populations, such as lower income and racial-ethnic minority households, are disproportionately impacted by disruptions in the power system. Moreover, the connected network qualities of the power system suggests an element of spatial vulnerabilities. However, little empirical evidence exists regarding the presence and extent of energy inequality. A main roadblock is the data collection process, in that outage data is often perishable and not found at granular spatial scales to allow the undertaking of a comprehensive analysis on impacts of power losses. Recognizing this important gap, this study collected and analyzed observational data related to the managed power outages during Winter Storm Uri (2021) and the hazard-induced outages during Hurricane Ida (2021). The research quantified the period of recovery at a granular spatial scale using an equitable-focused analysis to detect social and spatial inequalities through an exploratory lens. In extreme cases of power outage, census tracts of lower income and higher percentage of Hispanic population had longer median durations of recovery during Winter Storm Uri.  In the hazard-induced outages of Hurricane Ida, non-coastal zip codes with lower income had a 1.00-day longer median duration of recovery and higher percentage of Black population had a 2.00-day longer median duration of recovery while coastal zip codes with higher percentage of Black population had a 1.00-day longer median of recovery. Non-coastal regions had 63% greater spatial Gini values and 16% greater value in infrastructure inequality when compared to coastal regions. The managed power outages resulted in a 3% to 19% greater value of infrastructure inequality to the hazard-induced power outages. The findings provide evidence of pervasive social and spatial inequality in power outages during climate hazards and highlight the importance of integrating equity into the manner in which utility managers and emergency planners restore power outages.

## 1. Introduction

A combination of urbanization and climate change has contributed to the increasing severity, duration, and frequency of climate hazard events[1, 2]. Hazard events have transformed from rare, intense occurrences to frequent, inevitable disruptions which must be accounted for in the risk mitigation strategies of critical infrastructure[3, 4]. Critical infrastructure systems are highly integrated into the standard way of life and provide essential services to residents and community at large[5]. Unfortunately, these systems are increasingly susceptible to disruptions during hazard events[6-8]. The energy sector, in particular, has been severely impacted by extreme weather conditions including intense wind, heat, and freezing events[9]. Negative impacts on households are substantial as residents are deeply dependent on the multiple energy services to heat and cool their homes, cook and store fresh food, maintain a level of safety and security, power household appliances, and charge electronic devices. In addition, power systems are often interdependent with other critical infrastructure such as communication, water, and health services, worsening the impacts at a community scale[10, 11]. The uncertainties surrounding power disruptions places significant stresses on individual households and the extended community which must plan and endure for these inevitable outages.

Large-scale power outages can be hazard-induced or managed. Power systems are not only disrupted from direct physical damage but also from the imbalance in supply and demand[7, 12]. Hazard-induced outages can occur when hurricane winds down electrical lines, evidenced by the disruption of power during Hurricane Ida (2021) throughout the state of Louisiana[13], although outages were concentrated on the Louisiana coast in direct line of the hurricane path. In some affected areas, power was not restored until several weeks after the hurricane. In contrast, managed power outages are the protective measures taken when utility companies intentionally throttle supply. This is due an anticipated demand surge that would surpass supply. The process mitigates long-term damage to the power system [14]. At the network scale, extreme heat and freezing events can directly disrupt the transmission and supply lines, which are not ruggedized for operation in temperature extremes [15]. Electrical power providers preemptively shut down power to minimize electrical damages from downed power lines. Meanwhile, at the household level, energy demand remain high for heating and cooling systems to counteract extreme temperatures. An already weakened power system may be overexerted by the high demand of energy to the point of complete shutdown. Energy officials followed this strategy during Winter Storm Uri (2021) in Texas[16]. Rolling power blackouts cut power to neighborhoods and businesses with the attempted exception of critical facilities such as hospitals[17]. Rolling blackouts allow a set of residents to use the power for certain periods of time and then temporarily shut down their power to move power service to another set of residents[18]. Ideally, this practice allows all residents to evenly share the burden of periodic power outages throughout the hazard event, but this incorrectly assumes that residents will actually share the same energy burden.

Multiple knowledge gaps exist in the current research on energy inequality and power restoration. First, utility companies often restore power systems based on the extent of physical damage, number of outages, or total population, a method that may overlook how affected populations are differentially experiencing power losses[17, 19-21]. Recent policy and law discussions have brought attention to the universal accessibility of energy, but this discussion has not been reflected in the restoration of power outages [22-24]. Emerging research supports the equitable restoration of infrastructure disruptions [25-28], and by extension, restoration of the power system, for socially vulnerable populations [29-31]. Lower income and racial-ethnic minority populations, in particular, have reported disproportionate hardships and a diminished ability to mitigate power losses [32, 33]. Additional studies on Hurricane Maria in Puerto Rico and the wildfires in California highlighted that social vulnerability may complicate the restoration of power systems [31, 34]. The second knowledge gap is a failure to consider spatial inequalities in the restoration process. Power systems, by nature, are an interconnected and far-reaching network which could be susceptible to highly concentrated regions of impact [35, 36]. Third, models and decision frameworks for power outages primarily focus on the restoration of transmission lines [37], the creation of redundancy in links and nodes [38], and the perspectives from expert opinion [39, 40]. Though more models and frameworks have begun to include an equitable perspective[30, 41], there still lacks an understanding to the social and spatial vulnerabilities at a household perspective which can only be mitigated by analyzing empirical evidence of the intensity and duration of power outages.

Indeed, the availability of empirical data is quite limited. Power outage data is typically perishable and not available at granular spatial scales [42-44]. These limitations have hindered the ability to examine and understand the presence and extent of energy inequality in climate hazards. Recognizing this gap, in this research, we have collected and utilized observational data to calculate the percentage of outages at zip code and census tract scales, respectively, for two recent major events: the hazard-induced outages in the 2021 Hurricane Ida and the managed outages in the 2021 Winter Storm Uri. We analyzed these empirical data to quantify the extent of social and spatial inequalities and answer the following research questions: (1) to what extent power outage durations varied inequitably across different areas with different income and minority status?; (2) what was the extent of spatial inequality in the duration of power outages for each event?; (3) to what extent managed power outages led to greater spatial inequality compared with hazard-induced outages? In the analysis, first, we quantified the duration of service restoration for the two hazard events. Second, we determined the differential outage durations for low-income and minority populations within the two climate hazard events. Third, we compared the extent of spatial inequality between the hazard-induced and managed power outage events. The findings will bring attention into the equity gaps of power restoration, spotlight the importance of data transparency, and advance the understanding of mitigation strategies for energy inequality in climate hazard events.

## 2. Review on Relevant Literature

The literature review opens with the importance of researching and developing equitable infrastructure for normal and disaster conditions. It also discusses the role of decision makers and utility managers in equitable infrastructure. The review then narrows the scope to focus on disruptions in power systems. This includes the pervasive issue of energy inequalities and disproportionate impacts for power outages caused by disasters. The review closes with a discussion on the knowledge gaps in the studying power outages.

### 2.1 Importance of Equitable Infrastructure

Infrastructure plays a vital role in the recovery of households after a major disaster [6, 10]. People depend on infrastructure services like power, water, and transportation to return to their daily routines and maintain their well-being[5, 45]. However, not all community members have the same relationship with infrastructure. Certain populations may have less dependency, have greater accessibility, or can easily replace infrastructure services during disaster events. To maximize the recovery efforts of households after a disaster, there has been increasing interest in developing and cultivating equitable infrastructure[46-48]. For instance, the National Academies of Science identified "equitable and resilient infrastructure investments" as a priority for hazard mitigation research [49]. Equitable infrastructure addresses the systemic inequalities in communities to ensure everyone has access to the same opportunity and outcome of infrastructure services[49, 50]. Concurrently, resilient infrastructure ensures that infrastructure systems can recover to a level of functionality within a specified timeframe after the disaster[49, 51]. These two components ensure that all members in the community are able to recover after a disaster through accessibility and functionality of infrastructure services.

Decision makers and utility managers are influential to the managing equitable infrastructure since they play a key role in ensuring households receive adequate infrastructure services[5, 52]. However, the literature on equitable infrastructure reveals that equitable and social elements of infrastructure have not been properly considered. In his discussion of crises on the built-environment, Castaño-Rosa et. al [48] argues for the importance of using social interventions to achieve inclusive infrastructure management. Hendricks and Van Zandt [25] also state that the "built environment must be explored with a progressive lens that views physical infrastructure as an extension of social circumstances." They specifically mention how certain vulnerable communities such as low-income and racial-ethnic minority households could be unequally managed and protected in both daily and extreme events. Furthermore, Birkmann et. al [26] emphasizes that greater systematic knowledge of governance and human vulnerability is needed in infrastructure planning.

## 2.2 Disproportionate Impacts of Power Outages

Energy inequality is a significant issue in both normal and disaster conditions, and it has been debated from a human rights approach and within policy development [22-24]. Energy supply conditions, home energy efficiency, and the income affordability can all contribute to energy inequality[53]. Research in equitable infrastructure has also captured instances of energy inequality given in the form of disproportionate impacts of power disruptions. Through household survey data, Coleman et. al found repeated instances of increased hardship and longer disruption in low income and minority households from power outages caused by Hurricane Harvey, Hurricane Michael, and Hurricane Florence [32, 33]. Mitsova et. al [34] examined the differences in electric power outages and restoration rates in Hurricane Irma, and they showed a spatial dependency gap for lower income and less employed counties. In addition, Gargani [54] captured recovery through the energy production of the French territories impacted Hurricane Irma, and the indicators found that wealthier territories had comparatively quicker recovery rates.

Several studies have also investigated energy inequalities associated by Hurricane Maria, a powerful storm that caused months of power outages in Puerto Rico. Investigated by Garcia-Lopez [55], the impacts on critical infrastructure and resource systems including water, food, and energy were magnified by poor planning of federal institutions and decades of mismanagement. The wealthiest neighborhood in San Juan also had the quickest restoration time. In a related study, socially vulnerable populations were also less likely to be prioritized during disaster relief efforts based on crew deployments to restore power lines [56]. Satellite-based data revealed disproportionate duration of electricity outages in rural municipalities and lower income households [31]. Thus, social vulnerabilities are associated with increased impacts of outage which must be considered in the recovery process. Power restoration models have already begun to include aspects of social vulnerability[41, 57], but further research is needed to understand the nuances relationship between disaster impact, infrastructure disruption, and social vulnerability.

## 2.3 Research Gaps in the Energy Inequality

Despite advances in acknowledging equitable infrastructure[49], collecting relevant data through social media and satellite sources, [58-61] and integrating social vulnerabilities in power restoration models[41, 57, 62], there remain key research gaps in studying energy inequality in disasters. First, greater data transparency about the restoration of power is needed between utility companies and resident customers. Late reporting and inadequate information can lead to increased feelings of anxiety, worry, and discomfort as affected communities are left wondering about an essential service[63]. Second, power outage data is often perishable which means there is a lack of empirical data to support and inform decisions about power restoration [42]. Third, although certain power outage events report data, there is a lack of data at a granular spatial scale such as zip code or census tract [42]. Because of this, it can be difficult to connect the impacts of outages to the demographics of households or detect areas of spatial inequalities. Coarser-scale analysis can also overlook social vulnerabilities prominent in inner-cities due to the law of averages [64, 65]. In essence, the research study addresses these knowledge gaps by examining the outages during Hurricane Ida and Winter Storm Uri using observational outage data at a granular spatial scale. The study focuses on the potential disparities in the period of recovery to hazard-induced and managed power outages for low-income and racial-ethnic minority households. This research study contributes to the field of equitable infrastructure by examining the social and spatial impacts of an infrastructure system disruption. In particular, managed power outages are directly influenced by decision makers and utility managers, and thus, the research hopes the findings will highlight the importance of an equitable perspective in the restoration of power.

## 3. Background on Hazard Events

The research examined two recent weather events which significantly impacted the United States. The researchers aimed to select a weather event which represented managed power outages (Winter Storm Uri) hazard-induced outages (Hurricane Ida). This distinction is important as managed power outages can have direct implications in the decision-makers' response to outages while hazard-induced outages can reveal the restoration practices after extreme hazard exposure.

## 3.1 Managed Power Outage: Winter Storm Uri

Winter Storm Uri (WU) swept through a large swath of Central and Eastern United States between February 13 through 17, 2021. As the storm moved from the Pacific Northwest to the southern United States, it brought freezing temperatures with some areas experiencing new temperature lows and snowfall records [66].  Texas, one of states most affected by the storm, suffered an extended duration of power outages with certain areas reaching more than 3 consecutive days of outages against winter temperatures[67]. The harsh cold temperatures brought on by Winter Storm Uri resulted in a higher-than-normal energy demand by customers who mostly required it for heating their homes along with other essential services. Unable to sustain this unprecedented demand, power grids started failing. To manage the strain, the Southwest Power Pool (SPP) and the Electric Reliability Council of Texas (ERCOT) starting using rolling blackouts [68]. At some point, nearly 4.3 million customers were without power in Texas [69]. The use of rolling black outs, lasting power outages, and ice accretion caused disruption of water systems and water line ruptures with major affected areas, such as Harris County, being placed under boil-water advisories. This proved to be futile since, without power, people were unable to boil water to get rid of water pathogens and contaminants [68]. Winter Storm Uri cost the state approximately between $195 to $295 billion dollars in damage and the resulting increase in energy demand lead to thousands of additional charges [66, 70]. In the most extreme cases, the storm significantly impacted the well-being and health of people and eventually led to a number of fatalities. People reported cold stress symptoms including shivering, muscle stiffness, and loss of consciousness [15]. Reports placed the official death toll at 246 deaths in Texas, most which are attributed to from hypothermia due to dropping temperatures in homes and several fatalities due to carbon monoxide poisoning in an attempt to keep warm [71]. However, certain estimates place the death toll closer to 700 people, and specifically recognizing vulnerable populations with chronic conditions who were unable to withstand the freezing temperatures or were reliant on power for their medical devices [72]. Ultimately, the storm demonstrated the unpreparedness of the operation and management of the power grid to handle the severe shift in weather [68].

## 3.2 Hazard-Induced Outages: Hurricane Ida

Hurricane Ida (HI) first made landfall on August 29, 2021 near Port Fourchon, Louisiana, as a Category 4 hurricane coming in at a sustained speed of 150 mph [13, 73]. Louisiana, specifically the coastal region, suffered the brunt of the storm's damage. Power supplier Entergy, specifically, reported the greatest reports of outages as the energy company powered the majority of the affected areas. After Hurricane Ida passed through Louisiana, "more than 22,000 power poles, nearly 26,000 spans of wire, and more than 5,000 transformers [were] knocked over, broken or destroyed" [73]. As a result of the power outages, even more problems emerged, such as lack of power for critical facilities in hospitals and difficult living conditions in the summer heat with no air conditioning. Hurricane Ida was one of the deadliest hurricanes to impact the United States, causing significant destruction and greatly impacting many lives in its pathway. It caused an estimated $75 billion of damage, making it the fourth costliest Atlantic hurricane to make landfall in the United States [74]. With its deadly storm surges and intense rainfall and wind damage, weeks of power outages ensued in the hurricane's aftermath, and 91 human fatalities were recorded in Louisiana [75]. The structural levees held yet failed to protect all suburban areas, highlighting the failure to account for the intensification of hazard events and the ultimate inevitability of current and future hazard impacts [76].

## 4. Methods

The conceptual figure (Fig. 1) outlines the research steps. Outage data is collected and processed which is explained in detail in the following paragraphs. The methodology consists of six major parts. First, data processing of the telemetry-based data is a proxy measurement of outage data for Winter Storm Uri. Second, data collecting and imputation of real-time outage data from power company website for Hurricane Ida. Third, recovery of the outage data is calculated at a threshold of 10% disruption. Fourth, statistical methods analyze outage disruption in different demographic groups . Fifth, spatial Gini is evaluated to determine instances of spatial disparity. Sixth, infrastructure inequality index is calculated to understand the disparity of the power system. For Winter Storm Uri, one temporal cluster represents daytime outages between 8:00am to 8:00pm in Harris County. For Hurricane Ida, two spatial clusters, one cluster representing the coastal zip codes and one cluster representing non-coastal zip codes of Louisiana, are analyzed to distinguish between coastal influence and hazard impacts.

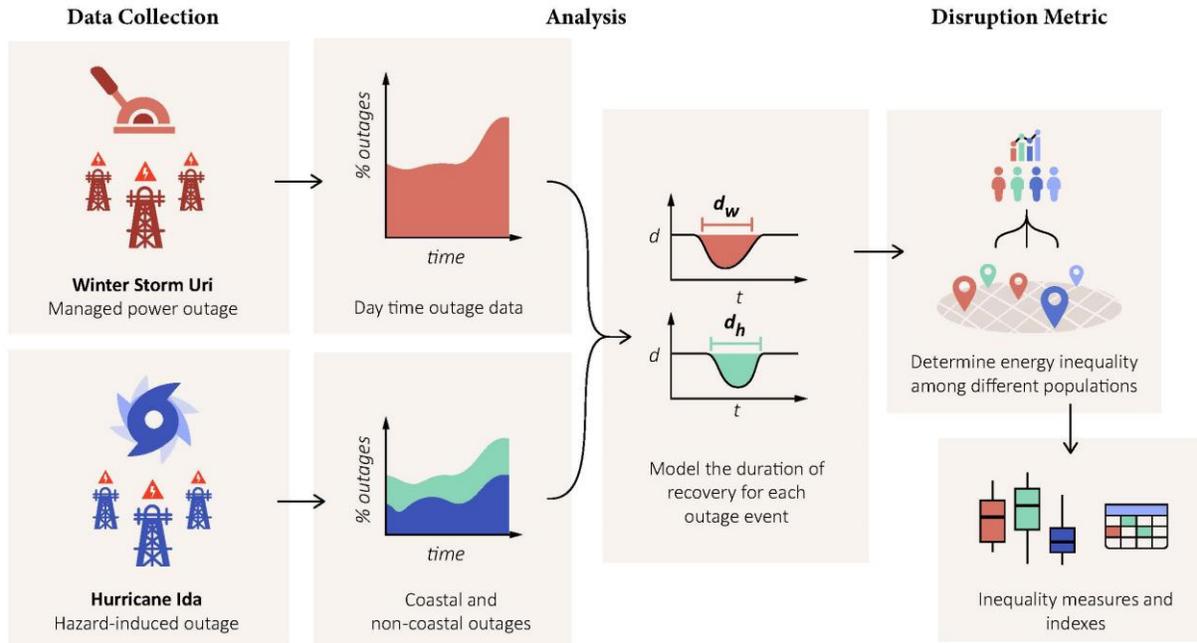

**Fig. 1.** Methods framework for the analysis of power outages during Winter Storm Uri and Hurricane Ida. The research team collected and processed observational data to quantify the duration of recovery. Analysis included descriptive and spatial statistics to determine energy inequality into inequality measures.

## 4.1 Telemetry Data Collection and Processing for Winter Storm Uri

Despite the concerted efforts of the research team, we were unable to procure direct and real-time outage information of Winter Storm Uri. Therefore, we quantified the power outage using telemetry-based population activity data through Mapbox. Mapbox, a business supplying mapping and location datasets [77], collaborated with the research team to provide aggregated raw cell phone information by time span and geographic unit. Mapbox collects the cell phone location from applications that use the Mapbox Software Development Kit (SDK). Two anonymization processes protected users' privacy. First, counts of target geographic areas were dropped if the counts fell below the minimum requirement. Second, slight random noise was applied to the existing counts. The spatial resolution of the aggregated data is about 100 by 100 meters; the temporal resolution is 4 hours. Thus, this raw point data is referred to as activity index. In essence, telemetry-based population activity data is able to capture fluctuations in cell-phone usage. The principle is that since power and communication services are often interdependent systems, telemetry-based data provides a reliable proxy measurement of outages [29]. For instance, telemetry-based population activity would decrease during blackouts which would provide a proxy of the extent and duration of power outages. Mapbox has been previously used in other studies to measure, by proxy, other forms of impact and disruption in disaster events which further supports its validity in the usage of this research study [78-82]. To obtain the proper outage information at the census tract, we converted the raw data, or activity index, to activity density. As discussed by Lee et.al. [29] we calculated the activity density within our desired geographic unit of census tracts for Harris County from February 1 through 28, 2021 (equation (1)).

$$Da(ct,t) = \sqrt{\frac{1}{N}\sum_{u=1}^{N} A_{u,t}^2} \qquad (1)$$

where $Da(ct,t)$ is the activity density *(Da)* in census tract *(ct)* at time *(t)*, $A_{u,t}$ is the activity index of the geographic unit *(u)* at time *(t)*, and $N$ is the number of geographic units within the census tract.

To ensure appropriate activity density from each census tract, the telemetry data spans from 8:00am to 8:00pm. In addition, the research team had both direct power outage data (further explained in the section below) and telemetry data for Hurricane Ida. Analysis between the power outage data and telemetry data showed a strong correlation, further supporting the proxy measurement of telemetry data for the outages in Winter Storm Uri. The respective analysis and figures are found in the Supplemental Information. Thus, moving forward, the research will refer to disruptions in the telemetry data as outage data for Winter Storm Uri.

## 4.2 Power Outage Collection and Processing for Hurricane Ida

Second, power outage data were collected in real-time from PowerOutage.US and the Entergy power websites during the storm event [83, 84] . The research focused on Entergy outages because this company was the main power provider to areas affected by Hurricane Ida in Louisiana. According to PowerOutage.US, Entergy provides energy to 1,275,873 residents in the state. To collect power outages data, we recorded the total number of affected customers and the last updated time, August 29 through November 23, 2021. From August 29 through September 3, 2021, data were collected each hour between 8:00 am till 12:00 am. From September 4 through October 25, 2021, power outages were recorded at three-hour intervals. From October 26 through November 23, 2021, power outages were only taken at 10:00 am, 4:00 pm, and 9:00 pm. The schedule was changed due to resulted from the decrease in noticeable number of outages.

We determined the percentage of outages at each zip code by dividing the number of affected customers, which was obtained from the Entergy website, by the total population at each zip code. It is important to note that Entergy makes no distinction on the outage website between business and residential customers. Since the researchers were unable to obtain the number of total customers per zip code, we normalized the outage data by the total population. This was to ensure that a greater population within a zip code was not erroneously inferred as a greater number of outages.

In addition, we cross-validated the power data from the Entergy website with the percentage of affected customers using county level data from PowerOutage.US. The county-level data from PowerOutage.US corroborated the Entergy data, so when one website showed a change in customers affected, the other showed a similar pattern on the same day. Note that in Louisiana, parishes are analogous to counties in other states. Since Entergy reported zip codes with at least one affected customer, we merged the Entergy data with all zip codes of parishes of interest. We selected only parishes where Entergy supplied at least 90% of total accounted customers, according to PowerOutage.US. To account for differences in coastal influence and potential impacts of hazards, we divided the affected regions into coastal and non-coastal zip codes based on the new coastal shape boundaries zone established in the Louisiana Revised Statutes Article 49 [85]. Coastal zip codes were drawn from the following parishes: Jefferson, St. Charles, St. Bernard, Plaquemines, St. John the Baptist, Orleans, Lafourche, Assumption, Terrebonne, Ascension, Iberville, Livingston, and Tangipahoa Parishes. Non-coastal zip codes were from the following counties: West Baton Rouge, East Baton Rouge, Pointe Coupee, East Feliciana, West Feliciana, St. Helena Parishes along with Ascension, Iberville, Livingston, Orleans, and Tangipahoa. Note that some counties contained both coastal and non-coastal areas.

For both datasets of Winter Storm Uri and Hurricane Ida, spatial units with more than 80% of NaN values were removed because the data could not confidently determine whether outages occurred. Missing data was filled with outage percentage values from the previous day. The time period for restoration analysis was February 14 , 2021 through February 24, 2021 for Winter Storm Uri and August 29 , 2021 through September 23 , 2021 for Hurricane Ida to best represent the period of major impact [13, 66].

### 4.3 Restoration of Power Services

Third, zip codes in Louisiana were considered to be "fully restored" when only 10% of households remained without power. Similarly, census tracts in Harris County were considered to "fully recovered" when the activity index change was less than 10%. This also meant that spatial units under the 10% threshold were not considered disruptive to account; these thresholds had to be consistent for at least two days. It is important for the research to acknowledge the ideal scenario that companies follow to restore power. The recovery prioritization process from energy providers begins with determining the areas with hazards such as downed lines/wires used for electrical distribution and damaged electrical equipment. Following this step, power companies initiate the repair process for lines and equipment to assist areas affected by the power outages. Then, energy providers restore power to critical facilities or critical community services which consist of hospitals, nursing homes, emergency response, utilities, and other significant services. Afterwards, power is restored to areas with the most affected customers in the shortest amount of time and later restore power to smaller affected areas. Energy providers such as SWEPCO also claim that "Residential customers are always given priority over business and industrial customers" [21]. As noted in the data collection and processing, the outage data collect cannot distinguish between residential and business outages; however, data was normalized by population. In addition, the 10% threshold is not a static measurement of recovery by power companies; however, the research determined this threshold based on a sensitivity analysis. The sensitivity analysis tested the duration of power restoration for 5%, 10%, and 20% thresholds (Supplemental Information).

### 4.4 Detecting Social Disparities

Fourth, the intent of the research was to address the extent, if any, of social disparities present during the outage events. The research will account for vulnerable populations delineated by income, racial, ethnic, and population differences. We collected sociodemographic information, including median household income, percentage of Black population, percentage of Hispanic population, and the total population, at the census tract and the zip code levels from the United States Census Bureau [80-82]. Data for the census tracts was collected from 2019 data while data for the zip codes was collected from 2020 data we examined the distribution of the duration of recovery, differences of populations impacted in descriptive statistics, and spatial representation of affected regions. In addition, median values and mean values of outage durations for the vulnerable populations, low-income and racial-ethnic minority populations, were compared through statistical tests at significance of $p<0.05$ which included correlations and ANOVA (Analysis of Variance).

**4.5 Spatial Gini Coefficient**

Fifth, to further understand the spatial inequality in the duration of recovery from outages, we first applied the spatial Gini coefficient [86]. The standard Gini index is traditionally used to demonstrate the wealth inequality of a region through a comparison of cumulative wealth, but the index can be used to measure the general inequality distribution of variables. The modified version of the index[86], called the spatial Gini decomposition (equation (2)) considers the spatial adjacency in a decomposition of the set of differences between "nearby" observations and the set of differences of "distant" observations.

$$G = \frac{\sum_i \sum_j w_{i,j}|x_i - x_j|}{2n^2 \underline{x}} + \frac{\sum_i \sum_j (1 - w_{i,j})|x_i - x_j|}{2n^2 \underline{x}} \tag{2}$$

where $w_{i,j}$ is a value one when $w_i$ and $w_j$ are neighbors and is zero otherwise.

The first term is the component among neighbors which describes the distances between nearby observations and the second term is the component among non-neighbors. According to Rey and Smith [86], spatial Gini is the first component that describes the differences between nearby neighbors. It tests whether the distribution of the components are different from those randomly distributed across a geographic space. In principle, inequality will be driven by disparities between distant places. In this way, spatial Gini index is much more sensitive to inequalities by considering pairwise relationships. Application of the spatial Gini can also produce pseudo p-values through assigned permutations which assumes a null hypothesis that values are randomly distributed geographically (equation (2)).

**4.6 Infrastructure Inequality Index**

Sixth, we also applied the more recently proposed infrastructure inequality index developed by Pandey, et. al [87]. To use the equation in our work for inequality of power outage durations, we normalized the outage durations based on a maximum-minimum value normalization for each event and then modified it as shown in Equation 3a. As shown in Equation 3b, the formula can translate outage durations into inequality in infrastructure provisions. In the equation from [87], μ is the mean of the extent of outages in a spatial area, where the mean should be a value between 0 and 1. In the equation, greater outage inequality would be closer to 1, vice versa.

$$y_{new} = \left(\frac{y - y_{min}}{y_{max} - y_{min}}\right) \tag{3a}$$

$$I = \frac{\sigma}{\sqrt{\mu(1-\mu)}}; \ 0 < \mu < 1 \tag{3b}$$

where μ is the mean of infrastructure provision values ($y$) obtained based on power outage values and $\sigma$ is the standard deviation of infrastructure provision values ($y$)

**5. Results**

## 5.1 Statistical and spatial representation to the duration of recovery to outages

The density of computed values for the duration of recovery to outages are seen in Fig. 2 and Fig. 3 for Winter Storm Uri and Hurricane Ida, respectively. In Fig. 2, the managed power outages were not normally distributed at statistically significant levels (p <0.05) which suggests a degree of inequality. The majority of census tracts experienced 0 days of recovery, meaning no period of power outages, followed by 6 days of outage and 10 days of outage. The mean value of power outages was 4.23 days during the 10-day analysis period. In Fig. 3, the hazard-induced power outages were uniformly distributed for coastal outages but was not uniformly distributed for non-coastal outages. Given the analysis period for the storm, there is also a statistically significant difference (p <0.05) in the mean values for coastal outages (11.77 days) and the mean values for non-coastal outages (4.55 days).

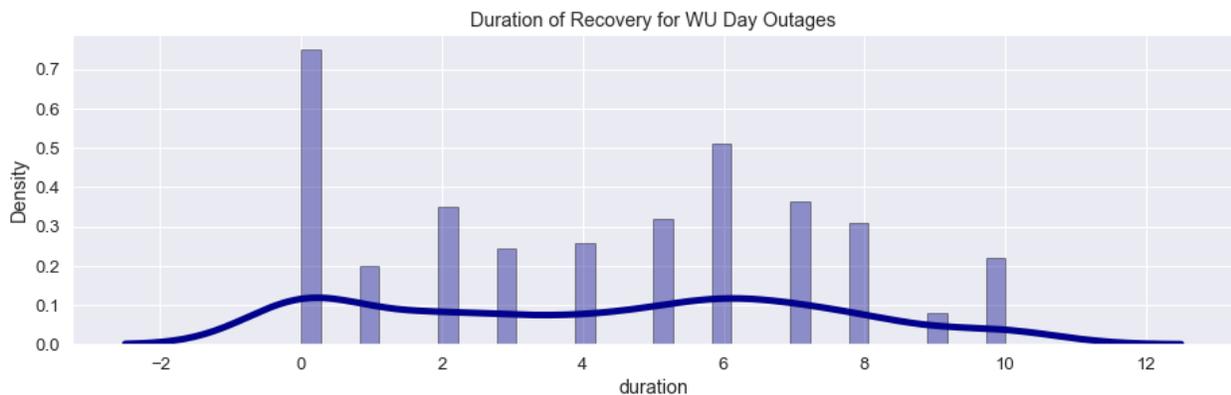

**Fig. 2.** Distribution of Managed Power Outages (in days) during the Day for Harris County. Duration of recovery was calculated at 10% threshold. The distribution curve was not normally distributed at a statistically significant level (p<0.05) and has slight peaks at 0, 6, and 10 days.

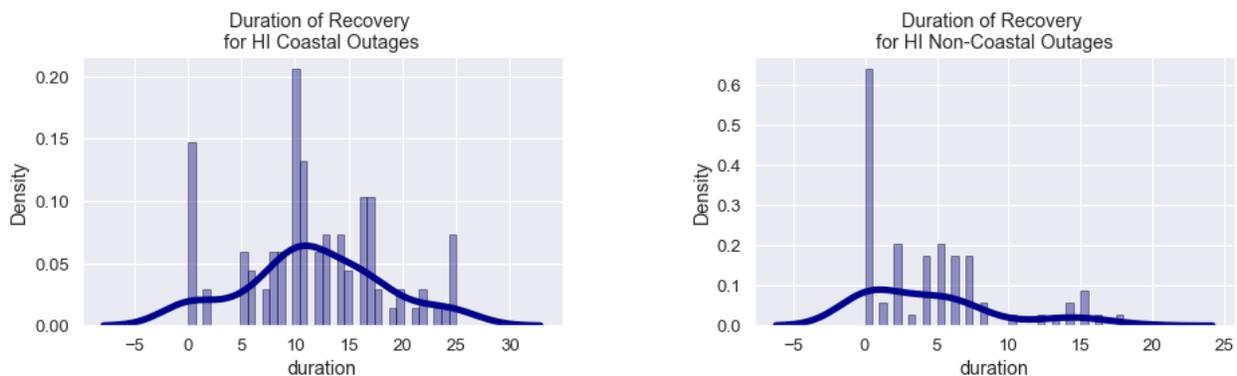

**Fig. 3.** Distribution of Hazard-Induced Power Outages (in days) for Coastal and Non-Coastal for Louisiana. Duration of recovery was calculated at 10% threshold. The distribution curve was normally distributed for coastal outages and not normally distributed for non-coastal outages at a statistically significant level (p <0.05). Coastal outages had slight peaks at 0, 10, and 25 days. Non-Coastal outages had slight peaks at 0 and 15 days.

Fig. 4 shows the spatial distribution of outage durations for the Winter Storm Uri and Hurricane Ida. In the case of Winter Storm Uri, the census tracts above the 75th percentile (greater than 7 days) of power outage typically lay in the central-east of Harris County while census tracts with 50th percentile (less than 1 day) of power outage lay in the outlying areas of the county. Regarding the coastal areas affected by Hurricane Ida, zip codes in the coastal areas above the 75th percentile (greater than 16 days) of outages were located in the southern coastal areas while zip codes below the 25th percentile (lesser than 8 days) were located in the northern coastal areas. Regarding the non-coastal areas affected by Hurricane Ida, zip codes above the 75th percentile (greater than 7 days) of outages were located in the eastern non-coastal areas while zip codes below the 25th percentile (at 0 days) were located in the western areas.

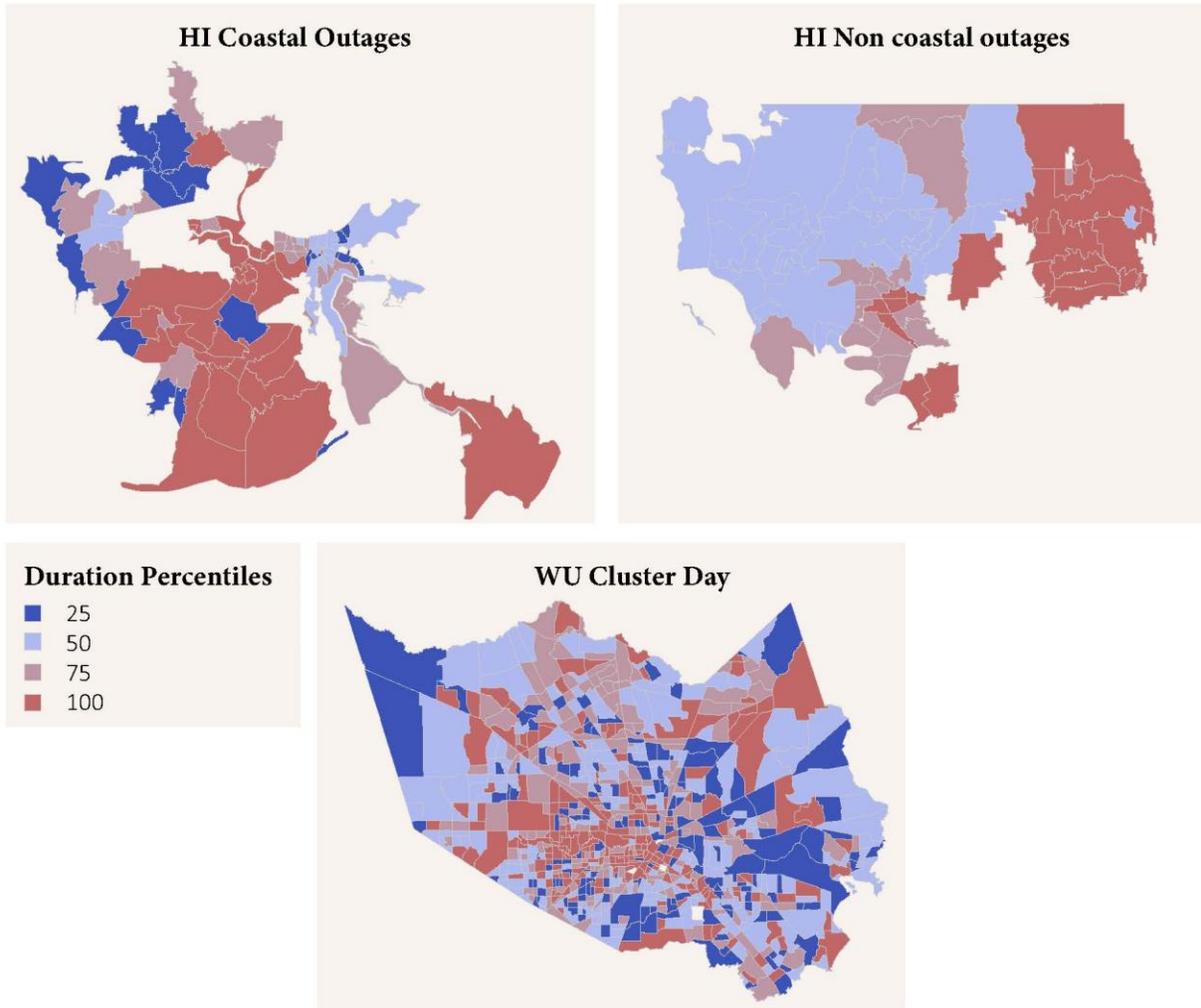

**Figure 4.** Distribution the hazard-induced power outages in coastal (upper left) and non-coastal (upper right) zip codes in Louisiana and the managed power outages for day outages in census tracts of Harris County. Zip codes above the 75th percentile were in the southern coastal areas and eastern non-coastal areas of Louisiana. Census tracts above the 75th percentile were mostly concentrated in the central-east center of Harris County.

## 5.2 Descriptive comparison of disparities in the duration of outages

We analyzed the power outage recovery duration for different demographic subpopulations based on median income, percentage of racial and ethnic minorities, and total populations based on percentile thresholds, as shown in Fig. 5 and Fig. 6. Overall, the linear correlations between the demographics and the duration of recovery was of little to no significance. Thus, we felt it appropriate to compare the two groups of lower and higher income, racial and ethnic minorities, and total population. The recovery duration to the managed power outages showed no significant difference among the social subpopulations when examining all the census tracts. However, the researchers decided to examine significantly impacted census tracts given the short duration but severe and unexpected nature of the winter storm. The analysis filtered census tracts with the highest percentage of outages. The percentage of extreme outages is defined as having a maximum percentage of outages of at least 70%. In Fig. 5, the box-and-whisker plots showed longer duration of recovery for census tracts with lower income and higher percentage of Hispanic residents.

Extremely impacted census tracts of lower income below the 50th percentile, or less than $41,657, had a 1.0-day greater median difference and 0.9-day greater mean difference (p<0.53) to the outage recovery duration when compared to census tracts of higher income. In addition to this comparison, we found that the lower income groups below the 25th percentile ($30,807) had a statistically significant mean difference in outage duration. There was a 2.00-day greater median difference and a 2.96-day greater mean difference to the power outage recovery duration (p < 0.05). Extremely impacted census tracts with higher percentages of Hispanic populations starting from above 29% (50th percentile) had a longer recovery duration when compared to census tracts of lower percentages with a 1.0-day median difference and a 1.56-day mean difference (p<.26). Specifically, having a greater than 80% percentage of Hispanic populations in the census tract led to a 2.50- day greater median difference to the power outage recovery duration while the 2.68-day greater mean difference was not statistically significant (p < .11), it does reveal a potential disparity in racial-ethnic communities. Of extremely impacted census tracts, lesser populated census tracts under the 50th percentile had a 1.00-day median longer recovery duration and .67-day mean difference, (p <.50) which shows that more populated census tracts had shorter duration of recovery.

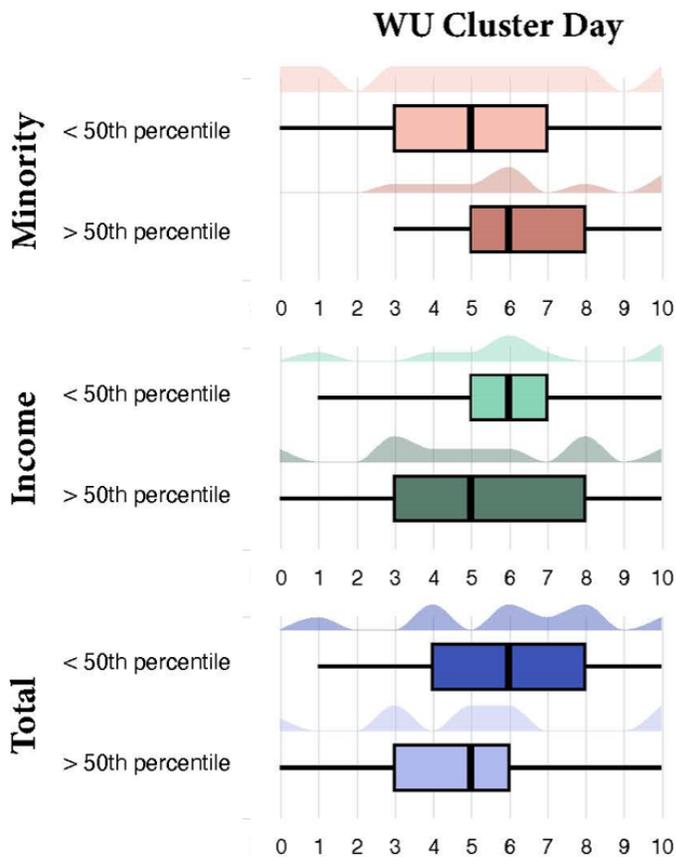

**Figure 5.** Descriptive statistics comparing the duration of recovery of the managed power outages in Harris County among different demographic characteristics. The analysis focused on census tracts with an extreme percentage of outages (70%). The box-and-whisker plots implies social disparities for census tracts with lower income at 50th percentile, higher percentage of Hispanic population at 50th percentile, and lesser populated at 50th percentile.

The coastal and non-coastal outages exhibited different results in terms of social inequalities. Even in cases where the results of the ANOVA means testing were not statistically significant, it is important to note the descriptive patterns of the demographic characteristics. As shown in Fig. 6, power outages in coastal areas had longer power outage recovery duration for zip codes for higher income (above 50th percentile or $51,831) while outages in non-coastal areas had longer power outage recovery duration for zip codes of lower income (below 50th percentile or $54,219). Coastal zip codes had 1.00-day median difference and 1.00-day mean difference (p <0.45) while non-coastal zip codes had a 1.00-day median difference and .40 mean difference (p < 0.75). Outages in both coastal and non-coastal areas had longer power outage recovery duration for zip codes of greater than 50th percentile of Black populations. Coastal zip codes had a 1.00-day median difference and .50 mean difference (p<.68) while non-coastal zip codes had a 2.00-day median difference and a .60 mean difference (p<.61). Overall, the results suggest that population size may have played a larger role in the duration of power outage recovery for coastal and non-coastal regions. Less populated zip codes had a longer power outage recovery duration for the coastal outage but more populated zip codes had a longer power outage recovery duration for non-coastal outages. Coastal zip codes of less population had a 4.00-day median difference and a 1.30-day mean difference (p<.34) while non-coastal zip codes of more population had a 5.00-day median difference and a 3.50-day mean difference at statistically significant values (p<0.002).

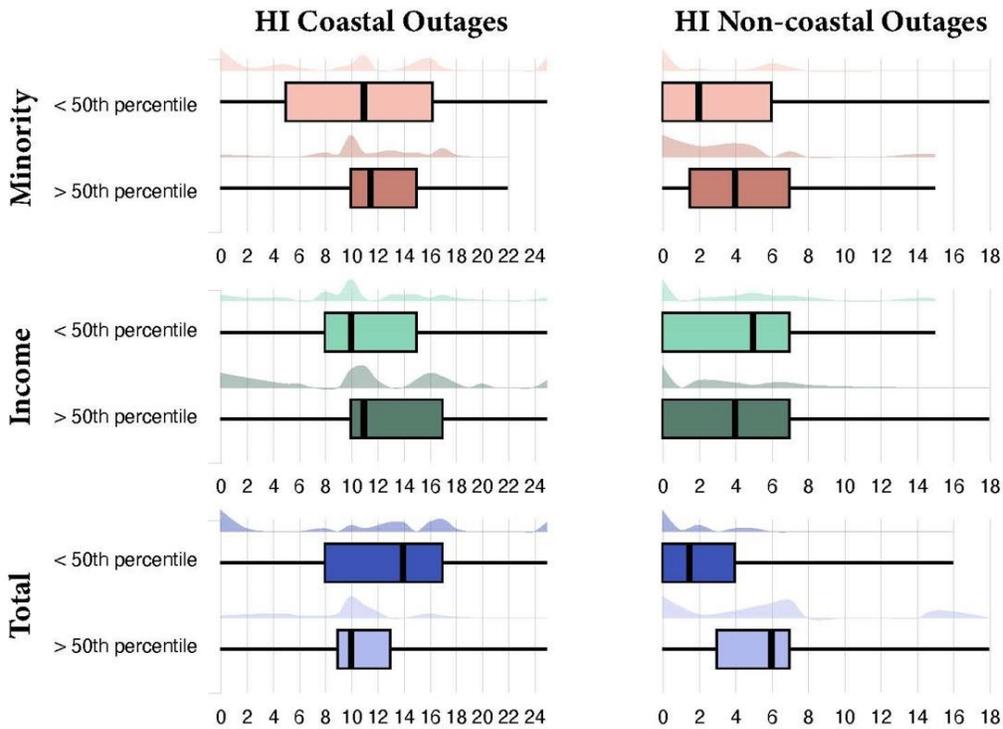

**Fig. 6.** Descriptive statistics comparing the duration of recovery of the hazard-induced power outages in Louisiana among different demographic characteristics. The analysis focused on all impacted census tracts. The box-and-whisker plots implies social disparities for census tracts with lower income and higher percentage of Black population at the 50[th] percentile.

**5.3 Spatial inequality indices for spatial heterogeneity of duration outages**

We examined the spatial inequalities in the duration of outages. First, we calculated the Spatial Gini coefficients and p-values as shown in Table 1. The statistical significance of the pseudo p-values (p < 0.05) indicates that spatial inequality between the neighboring pairs of the spatial units is different from the inequality of the pairs of spatial units which are not geographically proximate. In order, the greatest spatial Gini coefficients were the Hurricane Ida non-coastal outages, Winter Storm Uri day outages, and Hurricane Ida coastal outages. Though the coastal areas of Louisiana experienced the longest period of recovery from the power outages, these areas experienced the lowest spatial inequalities. This result means that the hazard exposure in conjunction with spatial heterogeneity must be considered when understanding the energy inequality for affected areas.

**Table 1.** Spatial Gini co-efficient values, p-values, and percent differences for duration of recovery of the managed and hazard-induced power outages

|  | Spatial Gini | % diff. to WU Day Outages | %diff. to HI Coastal Outages | %diff. to HI Non-Coastal Outages |
|---|---|---|---|---|
| **WU Day Outages** | **0.4214*** |  | 42.14 | -21.61 |
| **HI Coastal Outages** | **0.2727*** | -42.85 |  | -63.00 |
| **HI Non-Coastal Outages** | **0.5235*** | 21.61 | 63.00 |  |

*pseudo p-values are statistically significant at p< 0.01

To further compare the extent of spatial inequality between the managed outages and hazard-induced outages, we implemented the infrastructure inequality index [87]. Similar to the spatial Gini results, the coastal outages caused by Hurricane Ida, which experienced the most direct impact of the hurricane with the longest period of recovery, actually had the lowest infrastructure inequality. Specifically, the managed power outages caused by Winter Storm Uri had a 3.5% and 19.78% percent more spatial inequality to the hazard-induced outages in Louisiana. This result is significant in that it reveals, for the first time, that managed outages could be associated with greater inequality than hazard-induced outages. In other words, in the absence of equitable criteria for managed outage, such outages could cause a greater inequality in the extent of outages across a region compared with outages caused by hazards (such as wind damages to power transmission lines). The summary of the social and spatial vulnerabilities can be found in Table 3. We discuss the implications of these results in the next section.

**Table 2.** Infrastructure inequality index and percent difference calculations for duration of recovery of the managed and hazard-induced power outages

|  | Infrastructure Inequality Index | % diff. to WU Day Outages | % diff. to HI Coastal Outages | % diff. to HI Non-Coastal Outages |
|---|---|---|---|---|
| **WU Day Outages** | **0.638** |  | 19.78 | 3.49 |
| **HI Coastal Outages** | **0.523** | -19.78 |  | -16.31 |
| **HI Non-Coastal Outages** | **0.616** | -3.49 | 16.31 |  |

**Table 3.** Summary of the social and spatial vulnerabilities for the managed and hazard-induced power outages

|  | WU Day Outages | | HI Coastal Outages | | HI Non-Coastal Outages | |
|---|---|---|---|---|---|---|
| **Duration of Outages (Days) for Vulnerable Populations** | | | | | | |
|  | Median | Mean | Median | Mean | Median | Mean |
| <50th percentile of median income | 6.00 | 6.11 | 10.00 | 11.49 | 5.00 | 5.03 |
| >50th percentile of median income | 5.00 | 5.22 | 11.00 | 12.49 | 4.00 | 4.71 |
| <50th percentile of Black residents | | | 11.00 | 11.50 | 2.00 | 4.10 |
| >50th percentile of Black residents | | | 12.00 | 12.04 | 4.00 | 4.88 |
| <50th percentile of Hispanic residents | 5.00 | 4.88 | | | | |
| >50th percentile of Hispanic residents | 6.00 | 6.44 | | | | |
| <50th percentile of total population | 6.00 | 6.00 | 14.00 | 12.41 | 1.00 | 2.82 |
| >50th percentile of total population | 5.00 | 5.33 | 10.00 | 11.14 | 6.00 | 6.32 |

| **Values of Spatial Inequalities** | | | | | | |
|---|---|---|---|---|---|---|
| Spatial Gini | 0.42 | | 0.30 | | 0.54 | |
| Infrastructure Index | 0.64 | | 0.52 | | 0.62 | |

## 6. Discussion of Significant Findings

Recent climate hazard events have revealed the rising vulnerability of power systems, and the increasing intensity and duration of climate hazards will lead to even more disruptive power outage events. In facing future climate hazards, it is essential to examine the extent of inequality in outage durations to improve existing power restoration and equitably managing power outage schedules. This knowledge is limited; however, due in part to the perishable and hard-to-access nature of outage data. We addressed this important knowledge gap by using observational data at granular spatial scales from two major climate hazard events in the U.S. In doing so, our investigation uncovered the presence and extent of energy inequality during managed and hazard-induced power outages.

First, the results demonstrated the ability to quantify the duration of recovery for power systems using real-time observational data from direct power outage and telemetry-based data at a granular spatial unit. Research in telemetry-based data has captured aspects of community recovery including flood monitoring, return of evacuees, and business dependency, among others [78, 88-90] . However, to the best of the authors' knowledge, few studies have captured this fine resolution data to understand the impacts of energy inequality during climate hazard events. The research found that the effects of both managed and hazard-induced outages can last days beyond the initial impact point (climate hazard event). The median values of most affected zip codes of Hurricane Ida had 25 plus days of outage while the most affected census tracts of Winter Storm Uri had 10 plus days of outage. As shown in Fig. 4, census tracts in the 75th percentile of outages (greater than 7 days) lays in the central-east of Harris County. Coastal census tracts in the 75th percentile of outages (greater than 16 days) lay in the southern coastal areas and non-coastal census tracts in the 75th percentile of outages (greater than 7 days) were located in the eastern non-coastal areas. As shown in Fig. 2 and Fig. 3 for the Winter Storm Uri and Hurricane Ida, distribution curves also suggest potential inequality in the affected zip codes as trends were not normally distributed; instead, some had multiple peaks and heavy skewness. This finding showed the significant vulnerability of power infrastructure to climate hazards and the significant extent of outages experienced which surpass the capacity of power infrastructure owners, operators, and utilities to cope with the impacts. The standard practice prioritizes the restoration of areas with more outages and often more population. In contrast to standard practice, total population showed conflicting results in the duration of recovery to the power outages. As expected, more populated zip codes had a shorter duration of outage recovery in the most impacted areas of Harris County (Fig. 5) and coastal areas of Louisiana (Fig. 6) which shows that more populated areas could have been prioritized in the power restoration; however, more populated zip codes actually had a longer duration of outage recovery in the non-coastal areas of Louisiana (Fig. 6).

Second, the findings showed disparity in the outage durations for vulnerable social populations in both the managed and hazard-induced power outages. For Hurricane Ida in Louisiana, coastal zip codes with lower income along with non-coastal zip codes with lower income and higher percentage of Black populations had a longer duration of outage recovery (Fig. 6). Regarding the day outages caused by Winter Storm Uri in Harris County, these social disparities were more apparent for census tracts which were already experiencing high levels of impact. In these extreme cases, census tracts with lower income and higher percentage of Hispanic populations had longer duration of recovery (Fig. 5), which can further exacerbate systemic inequalities these social populations face during hazard events. Equitable infrastructure research has found that socially vulnerable populations can start with less capabilities to withstand power outages[5, 32, 33, 91]. This inherent inequity could be due to living in mobile homes and apartments, which are less structurally sound and capable of withstand the rigors of a storm, or not being able to afford power generators, which are too expensive for lower income families[92]. . Given the intense dependence that people have for energy, even a single day of difference could significantly impact households. Even in the instance of equal durations of outages, vulnerable populations may experience a disproportionate level of impact. Longer duration of outages for vulnerable populations compounds these disparities in impacts even further. Hence, it is essential for power infrastructure owners, operators, and utilities to establish transparent policies and criteria to integrate social equality into power restoration plans and actions.

Third, vulnerability can extend beyond the demographic make-up of the community and into the spatial components of the communities. Power systems, by nature of their network design, could be susceptible to spatial inequalities [35]. Statistically significant Gini indices indicated a layer of spatial heterogeneity in outage durations for both managed and hazard-induced outages. For hazard-induced outages, we found significant differences between the spatial inequalities of coastal and non-coastal zip codes. Though the coastal region had a longer duration of outage recovery, the non-coastal region had a 63% greater spatial Gini values (Table 2) and 16% greater values in the infrastructure inequality index when compared to coastal region (Table 3). Thus, spatial inequality cannot be completely explained based on coastal influences and direct impact to hazards. The infrastructure inequality index also revealed a significant difference between managed and hazard-induced outages. Managed power outage in Harris County impacted by Winter Storm Uri resulted in a 3% to 19% greater disparity in the infrastructure inequality index when compared to the hazard-induced outages in Hurricane Ida (Table 3). This finding is particularly noteworthy, since managed power outages are implemented based on assessment of power infrastructure managers and operators. Thus, the findings reveal both the limitations and potential improvement surrounding the current practices managed power outages. For instance, the standard practice is to address areas of larger population, the severity of impact, and the critical infrastructure and facilities of the community which all may unintentionally favor concentrated areas of the community over an equitable distribution. With the impacts of climate change (such as extreme heat and wildfires), more managed power outages are expected. The research underscores the importance of transparent communication and fair practices for implementing power outages. Power infrastructure managers, owners and operators should use proper metrics and criteria (such as those used in this study) to communicate to the public that their process for implementing managed outages was fair.

## 7. Conclusions and Future Research

In the discussion of energy inequality, multiple dimensions work in tandem to achieve equitable access to energy which are inclusive of all customers, show full transparency of data, and protection of human rights[22, 23]. Ostensibly, the current restoration practices fail to account for these dimensions. Despite the assumption that disasters are "natural" and thus equally affect a community, disaster research has continuously demonstrated that human decisions can lead to either the mitigation or exacerbation of disaster impacts. Systemic and inherent biases found in current restoration strategies burden certain populations and regions of a community. The findings of this study reveal the extent of social and spatial inequalities in power outage durations that must be carefully considered by energy policy makers and infrastructure managers in planning and implementing future power restoration actions.

Standard practices in restoration management may unintentionally isolate certain areas and populations of the community. Especially in managed-power outages, decision-makers can create equitable practices to ensure all members of the community recover. The research calls for more thoughtful consideration of the protection and management of power systems, those which consider not only the physical durability and the hazard exposure of the system but also the social and spatial structure receiving these vital services. As exhibited by the analysis, extremely impacted census tracts of lower income and higher Hispanic population had longer median duration of recovery for extreme cases of outages during Winter Storm Uri. Coastal zip codes and non-coastal had longer median duration of recovery for those with higher Black populations. Another example is the moderate to high values of Spatial Gini and infrastructure inequality index of the affected areas. Thus, the research provides evidence of social and spatial disparities which must be considered in both hazard-induced and managed power outages. In addition, future research could explore the morphology of the power grids across different cities along with extreme concentrations of critical facilities. This evaluation will uncover the role of grid morphology related to the spatial disparities of power outages across different cities and extreme events.

The current state of the outage data is not able to distinguish between business and residential losses; thus, more concentrated efforts should be made in collecting data at the business and household level. To properly quantify the disproportionate impacts of power losses, we must consider that businesses and households are often not affected equally. The impacts of power outages on residential customers are often complex as power can have direct and indirect disturbances on the physical, mental, and emotional state of residents[93]. In particular, small businesses in the food industry may be especially vulnerable to economic losses during power outage events [94]. By bringing attention to the potential social and spatial disparities, we anticipate that future studies can build upon these findings through the perspectives of businesses and households.

Future research in energy inequality could also consider the differential impacts of socially vulnerable populations in coastal and non-coastal areas. In summary, coastal areas with higher percentage of Black populations, higher income, and less populated had longer duration of recovery to the power outages while non-coastal areas with higher percentage of Black populations, lower income, and more populated had longer duration of recovery to the power outages. These differences in behavior were also found in inequality measures as non-coastal regions had 63% greater spatial Gini values and 16% greater value in infrastructure inequality when compared to coastal regions. Science-based research categorized coastal areas which were more susceptible to the impacts of coastal hazards [85]. This has direct implications in the distribution of limited resources and the development of local policies to protect, manage and restore coastal boundaries; however, there is little considerations of the different impacts on sociodemographic groups based on their place of residence. Thus, future research could further investigate the impacts of power outages on vulnerable populations between coastal and non-coastal areas.

Ultimately, these research avenues cannot be accomplished without the increased interest and investment in quantitative data of power outages. Decision-makers and researchers should work in tangent to collect, process, and examine fine-grained power outage data. Methodological frameworks and metrics, such as those in the paper, will bridge the gap between the theoretical understanding of outages to the practical application of real-time information.

## Acknowledgments

This material is based in part upon work supported by the National Science Foundation under Grant CMMI-1846069 (CAREER) and the support of the National Science Foundation Graduate Research Fellowship. The authors also would like to acknowledge the data support from Mapbox, PowerOutage.US, and Entergy power. Any opinions, findings, conclusions, or recommendations expressed in this material are those of the authors and do not necessarily reflect the views of the National Science Foundation and Mapbox.

## Author Contributions

All authors critically revised the manuscript, gave final approval for publication, and agree to be held accountable for the work performed therein. N.C. was the lead PhD student researcher and first author, who was responsible for supervising data collection, performed final analysis, and wrote the majority of the manuscript. A. E was responsible for guiding data collection and for developing the project. C.L provided guidance in the analysis of data as well as the supplemental information analysis of telemetry and outage data. . E.G assisted in data collection of the outage data and media research on the impacts of the disasters. P.K assisted in processing the outage data. A. M was the faculty advisor for the project and provided critical feedback on the project development and manuscript.

**Supplemental Information**
**Comparison of the direct power outage data and telemetry-based data for Hurricane Ida**

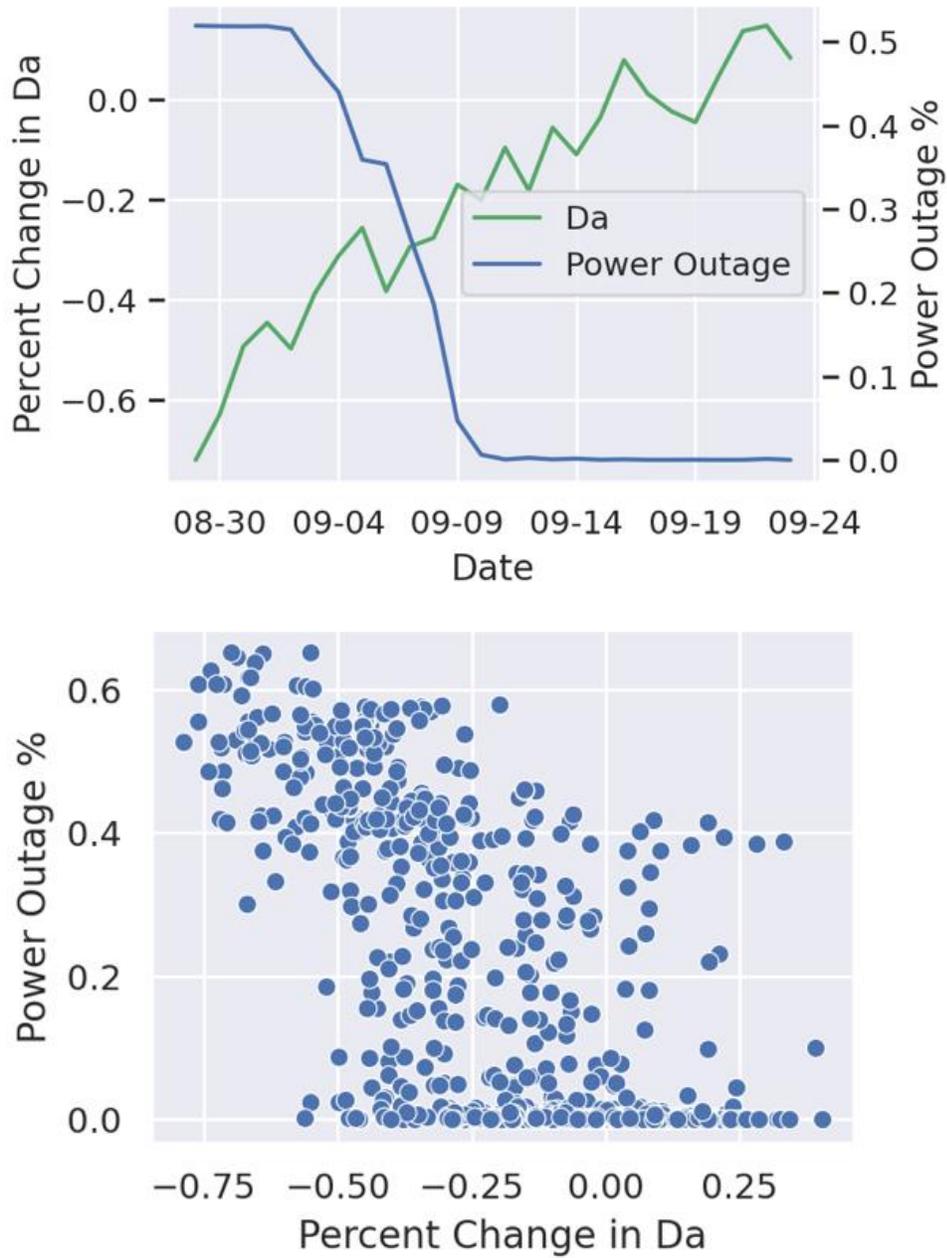

**Figure 1A .** The figures show the correlation between the percentage change in activity density of the zip code (Da) against the power outage percentage for Hurricane Ida. The top figure shows the overall relationship between August 30 through September 29 while the bottom figure shows a snapshot of the relationship.

**Sensitivity analysis to the varying thresholds to calculate the duration of recovery**

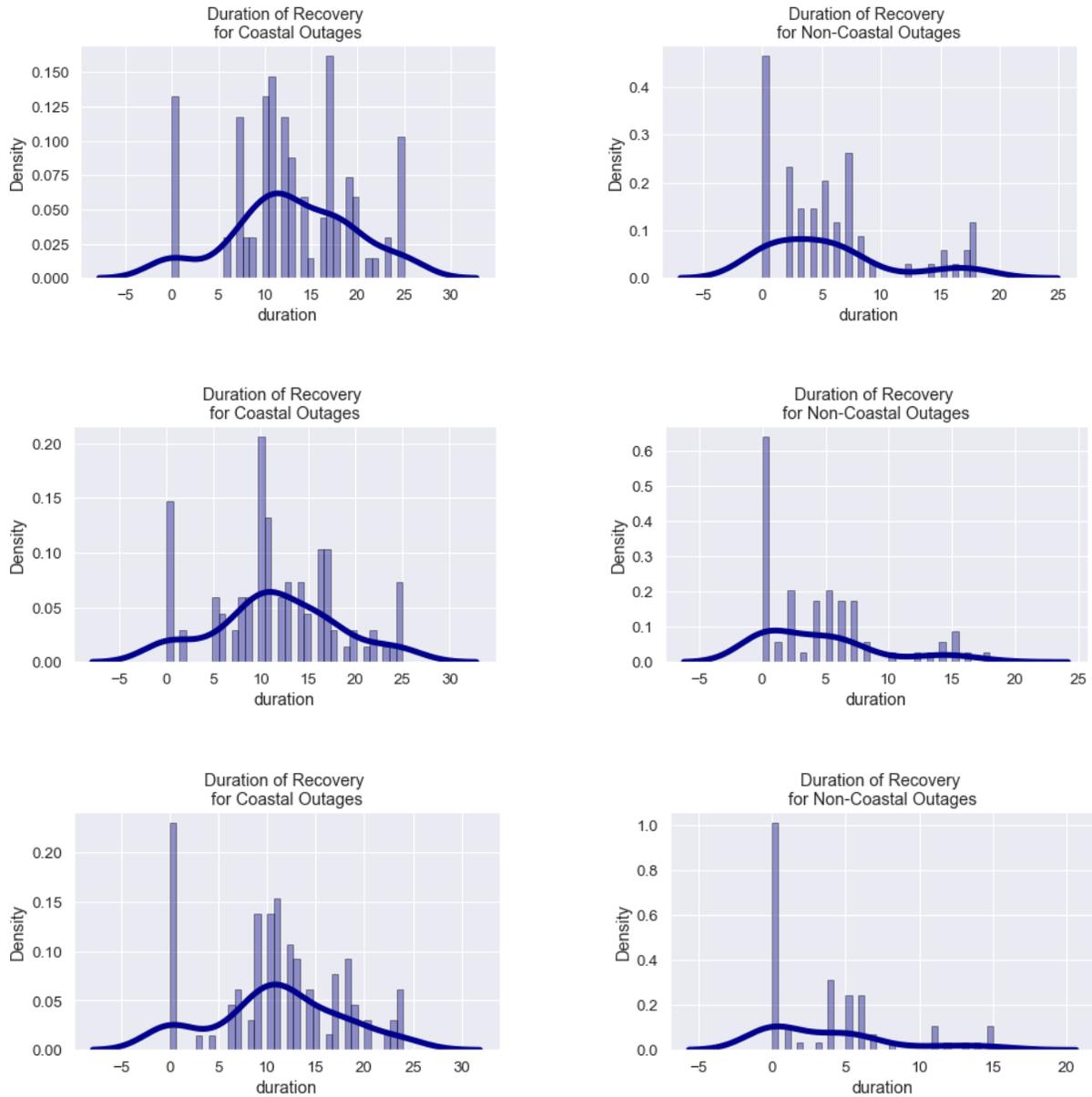

**Figure 2A.** The figures show the distribution of the duration of recovery for the coastal and non-coastal outages of Hurricane Ida. The values are listed from 20% threshold (top), 10% threshold (middle) and 5% threshold (bottom) of the percentage of outages.

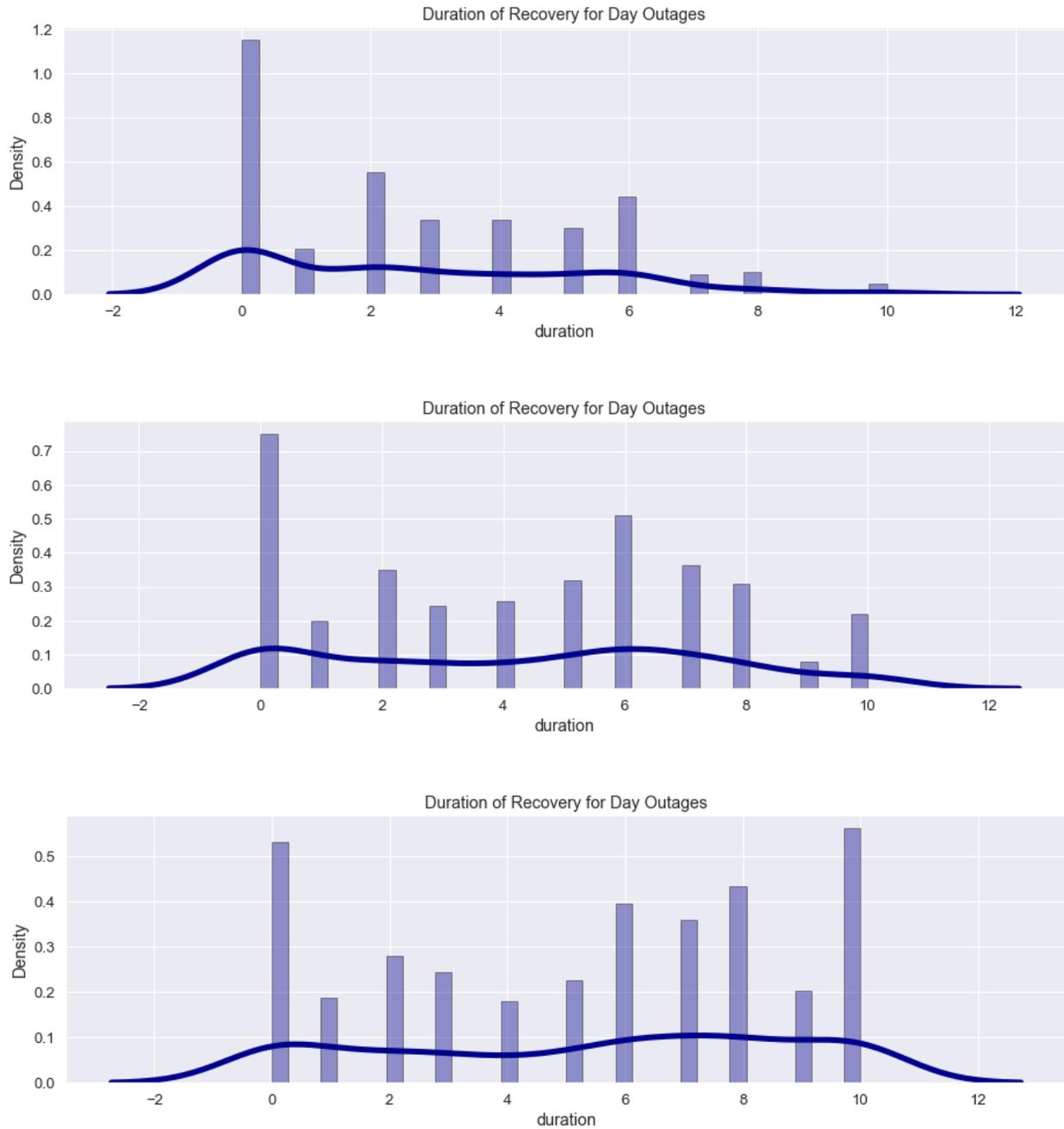

**Figure 3A.** The figures show the distribution of the duration of recovery for the day outages of Hurricane Harvey. The values are listed from 20% threshold (top), 10% threshold (middle) and 5% threshold (bottom) of the percentage of outages.

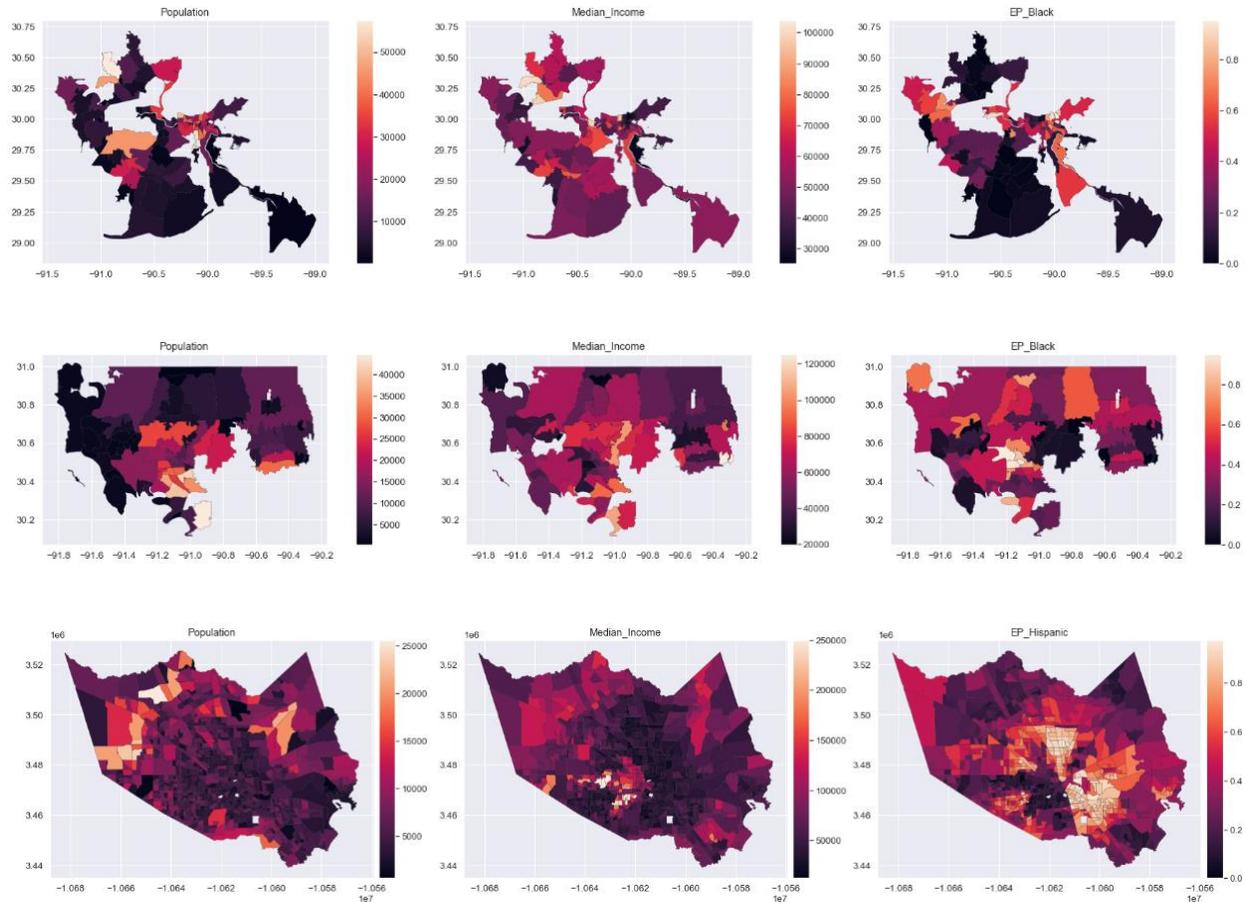

**Figure 4A.** The figures show the spatial distribution of the demographic features in the coastal zip codes of Louisiana (top), non-coastal zip codes of Louisiana (middle), and census tracts of Harris County.